\documentclass[a4paper]{article}
\usepackage{amssymb,amsmath,amsfonts}
\usepackage{fullpage}
\usepackage{multirow}
\usepackage{caption} 
\usepackage[latin1]{inputenc}
\begin{document}
\title{Asymptotic adjustments of Pearson residuals in exponential family nonlinear models}
\date{}
\author{Andréa V. Rocha$^{a,}$\footnote{E-mail: andrea.rocha@ci.ufpb.br}   and Alexandre B. Simas$^{b,}$\footnote{Corresponding author. E-mail: alexandre@mat.ufpb.br}\\\\
\centerline{\small{
$^a$Departamento de Computação Científica, 
Universidade Federal da Para\' iba,
}}\\
\centerline{\small{
$^b$Departamento de Matemática, 
Universidade Federal da Para\' iba,
}}}
\sloppy
\maketitle
\begin{abstract}
In this work we define a set of corrected Pearson residuals for continuous exponential family nonlinear models that have the same distribution as the true Pearson residuals up to order $\mathcal{O}(n^{-1})$, where $n$ is the sample size. Furthermore, we also introduce a new modification of the Pearson residuals, which we call PCA Pearson residuals, that are approximately uncorrelated. These PCA residuals are new even for the generalized linear models. The numerical results show that the PCA residuals are approximately normally distributed, thus improving previous results by Simas and Cordeiro (2009). These numerical results also show that the corrected Pearson residuals approximately follow the same distribution as the true residuals, which is a considerable improvement with respect to the Pearson residuals and also extends the previous work by Cordeiro and Simas (2009). 
\end{abstract}
\section{Introduction}
Residuals are essential to assess the fitting of a regression model, since it contains important information regarding 
the assumptions that underlie the statistical regression model being used. One of the main statistical tools currently in use
 by practitioners is the regression analysis. By looking at the definition of a regression model, we observe that they depend on many assumptions, 
so if one wants to apply a regression model to a particular data set, one must verify if the assumptions that underlie the statistical model hold. 
It is well-known that residuals contain important information on such assumptions, and therefore play an important role
in checking model adequacy. The use of residuals for assessing the adequacy of fitted regression models is
nowadays commonplace due to the widespread availability of statistical software, many of which are capable
of displaying residuals and diagnostic plots, at least for the more commonly used models.

In the seminal paper by Cox and Snell
(1968), they introduced residuals in a fairly general manner, thus being applicable to a wide class of models. 
They also discussed briefly the distribution of these residuals and the complementary aspect of transforming them, so that they have approximately the
same mean and variance as the ``true'' residuals. Loynes (1969) went further in this direction and provided
a transformation so that the resulting residuals has, approximately, the same distribution as the ``true''
residuals. Furthermore, Cox and Snell (1968) provided formulae for the first two moments (including the covariances) 
of the residuals up to the second-order. Nevertheless, it is interesting to note that, beyond special models, 
relatively little is known about asymptotic properties
of residuals in general regression models. There is a clear need to study second-order asymptotic properties of appropriate
residuals to be used for diagnostic purposes in nonlinear regression models. In this direction Simas and Cordeiro (2009) provided closed-form
expressions for the first two moments of Pearson residuals in exponential family nonlinear models (EFNLMs), up to the second-order. They thus used these expressions
to provide an adjusted residual having, up to the second-order, mean 0 and variance 1. This work extended the previous results by Cordeiro (2004), which provided similar expressions for Pearson residuals in generalized linear models (GLMs). Going on a different direction, Cordeiro and Simas (2009) applied Loynes' (1969) method to Pearson residuals in continuous GLMs and obtained a transformed Pearson residual, which they called the corrected Pearson residuals, having, up to the second-order, the same distribution as the ``true'' residual. One should observe that this last result was obtained only for the linear case, not holding for the exponential family nonlinear models.

The class of exponential family nonlinear models is a natural extension of
the well-known generalized linear models. 
This class of models is defined by a set of independent random variables
with a distribution in the exponential family and by a monotonic function that relates the mean
response to a nonlinear predictor involving covariates and unknown regression parameters, and has been introduced by Cordeiro and Paula (1987). 
The definition of such models includes a dispersion parameter: for example, the variance in normal
models. The book by Wei (1998) gives a comprehensive introduction to EFNLMs. The differential
geometric framework is presented for these models, and the geometric methods are widely used
in this book. The author also pays more attention to regression diagnostics and influence analysis.

Nelder and Wedderburn (1972) first identified and unified the theory for generalized linear models, 
including a general algorithm for computing maximum likelihood estimates (MLEs). Residuals in GLMs were first discussed by Pregibon (1981),
though ostensibly concerned with logistic regression models, Williams (1984, 1987) and Pierce and Schafer (1986). 
McCullagh and Nelder (1989) provided an excellent survey of GLMs, with
substantial attention to the definition of residuals. Residuals are used to identify discrepancies
between models and data,so it is natural to base residuals on the contributions made by individual
observations to measures of model fit. Pearson residuals are the most commonly used measure
of overall fit for GLMs and EFNLMs. They are defined as the signed square roots of the components of the
Pearson goodness-of-fit statistic by $R_i = (Y_i - \widehat{\mu}_i)/\widehat{V}_i^{1/2}$, where $\widehat{\mu}_i$ and
$\widehat{V}_i$ are, respectively, the fitted mean and variance function of $Y_i$. In this paper, 
we consider only Pearson residuals appropriate to our particular asymptotic aims when $n \to\infty$. 
Note that the Pearson residuals defined in Cordeiro (2004) are proportional to $\sqrt{\phi}$, although here we follow Simas and Cordeiro (2009) and Cordeiro and Simas (2009) and define it without the precision parameter $\phi$.

Our goal in this work is twofold. At first, we want to apply Loynes' (1969) method to the Pearson residuals in continuous EFNLMs to obtain transformed Pearson residuals having, up to the second-order, the same distribution as the ``true'' Pearson residuals, thus extending the previous work by Cordeiro and Simas (2009). We will follow Cordeiro and Simas (2009) and also call these residuals ``corrected Pearson residuals''. Secondly, we want to introduce a modification of the Pearson residuals in EFNLMs so that the resulting residuals do form a nearly independent set of random variables having, approximately, mean 0 and variance 1. More precisely, we want to provide a new set of residuals having, up to the second-order, mean 0, and identity covariance matrix. This new set of residuals are suitable to replace the adjusted Pearson residuals previously introduced by Simas and Cordeiro (2009), since, as the simulation results will show, they are better approximated by a standard normal distribution than the adjusted Pearson residuals introduced by Simas and Cordeiro (2009). Further, they are also suitable to identify violation of assumptions in the fitted model. We will call these residuals the PCA Pearson residuals, terminology introduced by Rocha and Simas (2016). The PCA Pearson residuals introduced in this work cannot be used to check for outlying observations, since they are obtained from a linear transformation of the ordinary Pearson residuals. Thus, in order to obtain a thorough description of the model, our suggestion is that both the PCA and corrected Pearson residuals should be used to assess a fitting of an exponential family nonlinear model, since the second is better to check the model assumptions, as will be seen in the simulation results, whereas the first can be used to check for outlying observations and also to verify the model assumptions.

To build a set of independent residuals, we apply the idea introduced in Rocha and Simas (2016) and use an idea reminiscent of principal component analysis. More precisely, we provide a new set of residuals that are ranked by their variance, and which are pairwise uncorrelated. The key ingredient is to work with their covariance matrix. Our results for the PCA Pearson residuals in EFNLMs provide an approximate result. Nevertheless, the results obtained by Rocha and Simas (2016) for the classical normal linear models are exact and works as paradigm for defining such residuals.

The remaining of the article unfolds as follows. In Section 2 we provide a brief review on exponential family nonlinear models. We obtain the corrected Pearson residuals for continuous EFNLMs in Section 3. Section 4 introduces the PCA Pearson residuals. In Section 5 we provide simulations results. Finally, in Section 6 we provide some concluding remarks.

\section{Exponential family nonlinear models}
In this section we provide a brief review on exponential family nonlinear models.
Let $Y_1,\ldots,Y_n$ be independent random variables, with each $Y_i$ having a density function in the linear exponential
family
\begin{equation}\label{dens}
\pi(y;\theta_i,\phi) = \exp\{\phi[y\theta_i-b(\theta_i)]+c(y,\phi)\},
\end{equation}
where $b(\cdot)$ and $c(\cdot,\cdot)$ are known appropriate functions. We assume that the precision parameter $\phi=\sigma^{-2}$ ($\sigma^2$ is 
called the dispersion parameter) is the same for all observations, although possibly unknown. If $Y$ is continuous, $\pi$ is assumed
to be a density with respect to the Lebesgue measure, while if $Y$ is discrete, $\pi$ is assumed to be a density with respect to the counting measure.
We have $E(Y_i) = \mu_i = db(\theta_i)/d\theta_i$ and var$(Y_i)=\phi^{-1}V_i$, where $V_i = d\mu_i/d\theta_i$ is the variance function.

The EFNLM is defined by equation \eqref{dens} and by the systematic component 
\begin{equation}\label{syst}
 g(\mu_i) = \eta_i = f(x_i;\beta),
\end{equation}
where $g(\cdot)$ is a known one-to-one differentiable link function, $x_i$ is a $q\times 1$ vector, $\beta=(\beta_1,\ldots,\beta_p)^T$
for $p<n$ a set of unknown parameters to be estimated, and $f(\cdot;\cdot)$ a nonlinear function in $\beta$ assumed to be continuously 
differentiable with respect to the vector $\beta$ such that the $n\times p$ derivative matrix of the nonlinear predictor, namely
$\widetilde{X} = \widetilde{X}(\beta) = \partial \eta/\partial \beta,$ has rank $p$ for all $\beta$. The $n\times p$ local model matrix $\widetilde{X}$
in general depends on the unknown parameter $\beta$. From the previous description it is easy to see that EFNLM extends both GLMs and
the nonlinear normal regression models. For GLMs, we have $p=q$, and the function $f$ assumes the form $f(x_i;\beta) = x_i^T\beta$,
where $x_i^T = (x_{i1},\ldots,x_{ip})$ is a vector of known variables associated with the $i$th observable response $y_i$. In many applications,
the regression function $f(\cdot,\cdot)$ has a linear component describing the experimental or observational conditions under which the 
observations were made.

We denote the log-likelihood for a given EFNLM by $l(\beta)$ and assume that equations \eqref{dens} and \eqref{syst} satisfy the
usual assumptions of large-sample likelihood theory (see, for instance, Cox and Hinkley, p. 107). The nonlinear predictors $x_1,\ldots,x_q$
are embedded in an infinite sequence of $q\times 1$ vectors that must satisfy these regularity conditions for the asymptotics to be valid.
Further, Lehmann and Casella (Chap. 6) showed that under these assumptions, the MLE $\hat{\beta}$ of the vector parameter $\beta$ has good
asymptotic properties, such as consistency, sufficiency, and normality. Conditions to assure weak consistency and asymptotic normality
of $\hat{\beta}$ in GLMs have previously been given by Fahrmeir and Kaufmann (1985).

The score function for $\beta$ is the $p\times 1$ vector $U(\beta) = \phi\widetilde{X}^TWP(y-\mu)$, where 
$(y-\mu) = (y_1-\mu_1,\ldots,y_n-\mu_n)^T$, $W = {\rm diag}(w_1,\ldots,w_n)$ is the diagonal matrix of weights $w_i = V_i^{-1}\mu_i'^2$,
$\mu_i'=\partial\mu_i/\partial\eta_i$, and from now on, primes indicate derivatives of the inverse link function with respect to the nonlinear
predictor $\eta$; and $P={\rm diag}(\mu_1'^{-1},\ldots,\mu_n'^{-1})$. The information matrix for $\beta$ is $K=E\{U(\beta)U(\beta)^T\}=\phi\widetilde{X}^TW\widetilde{X}$.
The MLE $\hat{\beta}$ can be obtained iteratively using standard reweighted least-squares method. The iteration is
$$\widetilde{X}^{(m)^T}W^{(m)}\widetilde{X}^{(m)}\beta^{(m+1)} = \widetilde{X}^{(m)^T}W^{(m)}t^{(m)},$$
where $t = \widetilde{X}\beta + P(y-\mu)$ is an adjusted dependent variable. The approximate covariance matrix of $\hat{\beta}$ is $\phi^{-1}(\widetilde{X}^TW\widetilde{X})^{-1}$.

\section{The distribution of Pearson residuals}
In this section, we will obtain the density of Pearson residuals to order $\mathcal{O}(n^{-1})$. We follow the approach of Loynes (1969). 
Since Loynes' approach only works on continuous random variables, we will assume that $Y$ is a continuous random variable, and so
the density in \eqref{dens} is with respect to the Lebesgue measure.

We will also adopt the following notation given in Cordeiro and Paula (1989):
$$(r)_i = \frac{\partial\eta_i}{\partial \beta_r},\quad (rt)_i = \frac{\partial^2\eta_i}{\partial\beta_r\partial\beta_t},$$
and so on.

Let $l_i$ be the log-likelihood contribution from $Y_i$. We have, from equation \eqref{dens}
$$l_i = \phi\{y_i\theta_i-b(\theta_i)\} + c(y_i,\phi),$$
and then, the $i$th term of the $r$th element of the score function is simply
$$U_r^{(i)} = \frac{\partial l_i}{\partial \beta_r} = \phi(Y_i-\mu_i)w_i^{1/2}V_i^{-1/2}(r)_i.$$

\subsection{Conditional moments of Pearson residuals}

Let $\varepsilon_i = (Y_i-\mu_i)V_i^{-1/2}$ be the true Pearson residual corresponding to the Pearson residual $R_i = (Y_i-\hat{\mu}_i)\hat{V}_i^{-1/2}$,
where $\hat{\mu}_i$ is the MLE of $\mu_i$. 

Suppose we write $R_i = \varepsilon_i + \delta_i$. We can write the conditional moments given $\varepsilon_i = x$ to order $\mathcal{O}(n^{-1})$ (Loynes, 1969)
as
$${\rm Cov}(\hat{\beta}_r,\hat{\beta}_s|\varepsilon_i=x) = -\kappa^{rs},$$
\begin{equation}\label{condbias}
b_s^{(i)}(x) = E(\hat{\beta}_s-\beta_s|\varepsilon_i=x) = B(\hat{\beta}_s) - \sum_{r=1}^p \kappa^{sr}U_r^{(i)}(x), 
\end{equation}
where $-\kappa^{sr}$ is the $(s,r)$th element of the inverse Fisher information matrix $K^{-1}$ for $\beta$, $B(\hat{\beta}_s)$ is the
$\mathcal{O}(n^{-1})$ bias of $\hat{\beta}_s$, and $U_r^{(i)}(x) = E(U_r^{(i)}|\varepsilon_i=x)$ is the conditioned score function. The
mean and variance of the asymptotic distribution of $\delta_i$, given $\varepsilon_i=x$ are, respectively, to order $\mathcal{O}(n^{-1})$,
\begin{equation}\label{thetai}
\theta_x^{(i)} = E(\delta_i|\varepsilon_i=x) = \sum_{r=1}^pH_r^{(i)}(x)b_r^{(i)}(x)-\frac{1}{2}\sum_{r,s=1}^p H_{rs}^{(i)}(x)\kappa^{rs}, 
\end{equation}
\begin{equation}\label{phii}
 \phi_x^{(i)^2} = {\rm Var}(\delta_i|\varepsilon_i=x) = -\sum_{r,s=1}^p H_r^{(i)}(x)H_s^{(i)}(x)\kappa^{rs},
\end{equation}
where $H_r^{(i)} = \partial\varepsilon_i/\partial\beta_r$, $H_{rs}^{(i)} = \partial^2\varepsilon_i/\partial\beta_r\partial\beta_s$,
$H_r^{(i)}(x) = E(H_r^{(i)}|\varepsilon_i=x)$ and $H_{rs}^{(i)}(x) = E(H_{rs}^{(i)}|\varepsilon_i=x)$. Let $V_i^{(m)}=\frac{d^mV_i}{d\mu_i^m}$
for $m=1,2$. We obtain by differentiation
$$H_r^{(i)} = \big\{-V^{-1/2}\mu_i' - \frac{1}{2}V_i^{-3/2}V_i^{(1)}\mu_i'(Y_i-\mu_i)\big\}(r)_i,$$
and
\begin{eqnarray*}
H_{rs}^{(i)} &=&  \Big\{-V_i^{-1/2}\mu_i''+V_i^{-3/2}V_i^{(1)}\mu_i'^2+\frac{3}{4}V_i^{-5/2}V_i^{(1)^2}\mu_i'^2(Y_i-\mu_i)\\
&&-\frac{1}{2}V_i^{-3/2}V_i^{(2)}\mu_i'^2(Y_i-\mu_i)-\frac{1}{2}V_i^{-3/2}V_i^{(1)}\mu_i'(Y_i-\mu_i)\Big\}(r)_i(s)_i\\
&+&\big\{-V^{-1/2}\mu_i' - \frac{1}{2}V_i^{-3/2}V_i^{(1)}\mu_i'(Y_i-\mu_i)\big\}(rs)_i.
\end{eqnarray*}

Conditioning on $\varepsilon_i=x$, we obtain $H_r^{(i)}(x) = e_i(x)(r)_i$ and $H_{rs}^{(i)}(x) = h_i(x)(r)_i(s)_i + e_i(x)(rs)_i$, where
\begin{equation}\label{efun}
e_i(x) = -V_i^{-1/2}\mu_i' - \frac{1}{2}V_i^{-1}V_i^{(1)}\mu_i'x, 
\end{equation}
and
\begin{equation}\label{hfun}
h_i(x) = -V_i^{-1/2}\mu_i''+V_i^{-3/2}V_i^{(1)}\mu_i'^2+\frac{1}{4}\{(3V_i^{-2}V_i^{(1)^2}-2V_i^{-1}V_i^{(2)})\mu_i'^2-2V_i^{-1}V_i^{(1)}\mu_i''\}x. 
\end{equation}

For canonical models ($\theta=\eta$), \eqref{efun} and \eqref{hfun} become 
$$e_i(x) = -V_i^{1/2}-\frac{V_i^{(1)}}{2}x\hbox{~and~}h_i(x) = \frac{1}{4}(V_i^{(1)^2}-2V_iV_i^{(2)})x.$$

Conditioning the score function $U_r^{(i)} = \phi V_i^{-1/2}w_i^{1/2}(Y_i-\mu_i)(r)_i$ on $\varepsilon_i=x$ yields $U_r^{(i)}(x) = \phi w_i^{1/2}(r)_i x$,
and then, using \eqref{condbias}, we obtain
$$b_s^{(i)}(x) = B(\hat{\beta}_s) + \phi w_i^{1/2}x\tau_s^TK^{-1}\widetilde{X}^T\gamma_i,$$
where $K^{-1} = \phi^{-1}(\widetilde{X}^TW\widetilde{X})^{-1}$, $W={\rm diag}\{w_i\}$ is a diagonal matrix of weights, $\tau_s$ is a $p$-vector
with one in the $s$th position and zeros elsewhere, and $\gamma_i$ is an $n$-vector with one in the $i$th position and zeros elsewhere.
Defining $M = \{m_{si}\} = (\widetilde{X}^TW\widetilde{X})^{-1}\widetilde{X}$, we have
$$b_s^{(i)}(x) = w_i^{1/2}m_{si}x + B(\hat{\beta}_s).$$

The  $\mathcal{O}(n^{-1})$ bias $B(\hat{\beta})$  of $\hat{\beta}$ was obtained by Paula (1992) and written as a matrix expression:
$$B(\hat{\beta}) = (\widetilde{X}^TW\widetilde{X})^{-1}\widetilde{X}^TW(\xi_1+\xi_2),$$
where $\xi_1 = -(2\phi)^{-1}Z_d W^{-1}F \boldsymbol{1}$, $\xi_2 = -(2\phi)^{-1}D\boldsymbol{1}$, $\boldsymbol{1}$ is an $n\times 1$ vector
of ones, $Z = \{z_{ij}\} = \widetilde{X}(\widetilde{X}^TW\widetilde{X})^{-1}\widetilde{X}^T$, $Z_d = {\rm diag}\{z_{ii}\}$, $D={\rm diag}\{d_1,\ldots,d_n\}$,
$d_i = tr\{\widetilde{X}_i(\widetilde{X}^TW\widetilde{X})^{-1}\}$, $\widetilde{X}_i$ is a $p\times p$ matrix with elements $\partial^2\eta_i/\partial\beta_r\partial\beta_s = (rs)_i$,
and $F = {\rm diag}\{V_i^{-1}\mu_i'\mu_i''\}$.

We are now in a position to calculate the quantities in \eqref{thetai}. The first term is given by
\begin{eqnarray*}
\sum_{r=1}^p H_r^{(i)}(x)b_r^{(i)}(x) &=& e_i(x)\left\{x w_i^{1/2}\sum_{r=1}^p m_{ri}(r)_i + \sum_{r=1}^p B(\hat{\beta}_r)(r)_i\right\}\\
&=& e_i(x)\{w_i^{1/2} z_{ii}x +\gamma_i^T \widetilde{X} B(\hat{\beta})\},
\end{eqnarray*}
whereas the second term is given by
$$-\frac{1}{2}\sum_{r,s=1}^p H_{rs}^{(i)}(x) \kappa^{rs} = \frac{z_{ii}}{2\phi} h_i(x) + \frac{d_i}{2\phi}e_i(x).$$

Therefore, the conditional mean $\theta_x^{(i)}$ from equation \eqref{thetai} is a second-degree polynomial in $x$ given by
\begin{equation}\label{thetaiform}
 \theta_x^{(i)} = \{w_i^{1/2}z_{ii}x + \gamma_i^T\widetilde{X}B(\hat{\beta})+\frac{d_i}{2\phi}\}e_i(x)+ \frac{z_{ii}}{2\phi}h_i(x).
\end{equation}

We now compute the conditional variance $\phi_x^{(i)^2}$. From \eqref{phii}, it follows that
\begin{equation}\label{phiiform}
\phi_x^{(i)^2} = \frac{z_{ii}}{\phi}e_i(x)^2. 
\end{equation}
Hence, $\phi_x^{(i)^2}$ is also a second-degree polynomial in $x$. 
\subsection{The corrected Pearson residuals}
Our goal in this section is to define, for each $i=1,\ldots,n$, a function $\rho_i(\cdot)$ in such a way that the corrected residual $R_i'$ given by
$R_i' = R_i + \rho_i(R_i)$ has the same distribution as $\varepsilon_i$ to order $\mathcal{O}(n^{-1})$. 

Let $f_{\varepsilon_i}$ be the density of the true Pearson residual. Loynes (1969) showed that $\rho_i$ is given, in terms of the conditional moments, as
\begin{equation}\label{formrho}
\rho_i(x) = -\theta_x^{(i)} + \frac{1}{2f_{\varepsilon_i}(x)} \frac{d\{f_{\varepsilon_i}(x)\phi_x^{(i)^2}\}}{dx}. 
\end{equation}

The density $f_{\varepsilon_i}$ is given by
$$f_{\varepsilon_i}(x) = \sqrt{V_i} \exp\left\{\phi\left[\sqrt{V_i}\theta_i x + \mu_i\theta_i - b(\theta_i)\right] + c\left(\sqrt{V_i}x+\mu_i,\phi\right)\right\},$$
whereas the density of $R_i$, namely $f_{R_i}$, is given by
$$f_{R_i} = f_{\varepsilon_i}(x) -\frac{d\{f_{\varepsilon_i}(x)\theta_x^{(i)}\}}{dx} + \frac{1}{2}\frac{d^2\{f_{\epsilon_i}(x)\phi_x^{(i)^2}\}}{dx^2},$$
where $\theta_x^{(i)}$ and $\phi_x^{(i)^2}$ are given in \eqref{thetaiform} and \eqref{phiiform}, respectively.

In Table \ref{tabela1} we provide the densities of the true Pearson residuals for the normal, gamma and inverse Gaussian distributions,
where $\Gamma(\cdot)$ is the gamma function.

\begin{table}[hbt]
\begin{center}
\caption{Densities of the true residuals for some distributions.}
\begin{tabular}{lcc}
\hline
Distribution&Density in \eqref{dens}& Density of the true residual $(f_\varepsilon(x))$\\
\hline
Normal&$\frac{1}{\sqrt{2\pi}\sigma}\exp\left\{-\frac{(x-\mu)^2}{2\sigma^2}\right\}$&$\frac{1}{\sqrt{2\pi}\sigma}\exp\left\{-\frac{x^2}{2\sigma^2}\right\},\qquad x\in\mathbb{R}$\\
Gamma&$\frac{(\phi x)^{\phi-1}\phi}{\Gamma(\phi)\mu^\phi}\exp\left\{-\frac{\phi x}{\mu}\right\}$&$\frac{\{\phi(1+x)\}^{\phi-1}\phi}{\Gamma(\phi)}\exp\left\{-\phi(1+x)\right\},\qquad x>-1$\\
Inverse Gaussian&$\frac{\sqrt{\phi}}{\sqrt{2\pi x^3}}\exp\left\{-\frac{\phi(x-\mu)^2}{2\mu^2x}\right\}$&$\left\{\frac{\phi}{2\pi(\mu^{1/2}+1)^3}\right\}^{\frac{1}{2}}\exp\left\{-\frac{\phi x^2}{2(\mu^{1/2}x+1)}\right\},\qquad x>-\frac{1}{\sqrt{\mu}}$\\
\hline
\end{tabular}
\label{tabela1}
\end{center}
\end{table}

Using equation \eqref{phiiform}, we obtain
\begin{equation}\label{formula1}
 \frac{1}{2f_{\varepsilon_i}(x)} \frac{d\{f_{\varepsilon_i}(x)\phi_x^{(i)^2}\}}{dx} = \frac{z_{ii}}{\phi} e_i(x) \frac{de_i(x)}{dx} + \frac{z_{ii}}{2\phi} e_i(x)^2\left\{\phi\sqrt{V_i}\theta_i + \frac{d}{dx}c(\sqrt{V_i}x+\mu_i,\phi)\right\}.
\end{equation}

Thus, from equations \eqref{formrho}, \eqref{formula1} and \eqref{thetaiform}, we obtain
\begin{eqnarray}
\rho_i(x) &=& e_i(x)\left\{ -\frac{1}{2\phi}V_i^{-1}V_i^{(1)}\mu_i'z_{ii} - \gamma_i^T\widetilde{X}B(\hat{\beta})-w_i^{1/2}z_{ii}x -\frac{d_i}{2\phi}\right\}\nonumber\\
&& - \frac{z_{ii}}{2\phi}h_i(x) + \frac{z_{ii}}{2\phi}e_i(x)^2\left\{\phi\sqrt{V_i}\theta_i + \frac{d}{dx}c(\sqrt{V_i}x+\mu_i,\phi)\right\}.\label{formularho}
\end{eqnarray}

The formula obtained in equation \eqref{formularho} is new and is one of the main results of the paper.
If we denote by $f_{R_i'}$ the density of the corrected residual. The function $\rho_i(\cdot)$ is such that $f_{R_i'} = f_{\varepsilon_i}$
to order $\mathcal{O}(n^{-1})$.

If the regression function $f(\cdot;\cdot)$ is linear, given by $f(x_i^T;\beta) = x_i^T\beta$, we have $d_i=0$, $B(\hat{\beta})$
reduces to the formula obtained by Cordeiro and McCullagh (1991), thus equation \eqref{formularho}
agrees with the formula obtained by Cordeiro and Simas (2009). 

In Table \ref{tabela2} we provide the values of $e_i(x)$ and $h_i(x)$ for the normal, gamma and inverse gaussian distributions. In Table
\ref{tabela3} we provide the values of $\mu',\mu''$ and $w$ for some link functions. Finally, in Table \ref{tabela4} we provide
the values of $\theta,V,w$ and $\frac{d}{dx}c(\sqrt{V}x+\mu,\phi)$ for the normal, gamma and inverse gaussian distributions.

\begin{table}[hbt]
\begin{center}
\caption{Values of $e_i(x)$ and $h_i(x)$ for the normal, gamma and inverse Gaussian distributions.}
\begin{tabular}{lcc}
\hline
Distribution&$e_i(x)$& $h_i(x)$\\
\hline
Normal&$-\mu_i'$&$-\mu_i''$\\
Gamma&$-\mu_i^{-1}\mu_i'-\mu_i^{-1}\mu_i'x$&$-\mu_i^{-1}\mu_i''+2\mu_i^{-2}\mu_i'^2-\mu_{-1}\mu_i''x+2\mu_i^{-2}\mu_i'^2x$\\
Inverse Gaussian&$-\mu_i^{-3/2}\mu_i'-\frac{3}{2}\mu_i^{-1}\mu_i'x$&$-\mu_i^{-3/2}\mu_i''+3\mu_i^{-5/2}\mu_i'^2+\frac{15}{4} \mu_i^{-2}\mu_i'^2x-\frac{3}{2}\mu_i^{-1}\mu_i''x$\\
\hline
\end{tabular}
\label{tabela2}
\end{center}
\end{table}

\begin{table}[hbt]
\begin{center}
\caption{Values of $\mu'$, $\mu''$ and $w$ for some link functions.}
\begin{tabular}{lcccc}
\hline
Link function&Formula&$\mu'$&$\mu''$&$w$\\
\hline
Linear&$\mu=\eta$&$1$&$0$&$V^{-1}$\\
Log&$\log(\mu)=\eta$&$\mu$&$\mu$&$\mu^2V^{-1}$\\
Reciprocal&$\mu^{-1}=\eta$&$-\mu^2$&$2\mu^3$&$\mu^4V^{-1}$\\
Inverse of the square&$\mu^{-2}=\eta$&$-\mu^3/2$&$3\mu^5/4$&$\mu^6V^{-1}/4$\\
\hline
\end{tabular}
\label{tabela3}
\end{center}
\end{table}

\begin{table}[hbt]
\begin{center}
\caption{Values of $\theta,V,w$ and $\frac{d}{dx}c(\sqrt{V}x+\mu,\phi)$ for the normal, gamma and inverse Gaussian distributions.}
\begin{tabular}{lcccc}
\hline
Distribution&$\theta$&$V$&$w$& $\frac{d}{dx}c(\sqrt{V_i}x+\mu_i,\phi)$\\
\hline
Normal&$\mu$&$1$&$\mu'^2$&$-(x+\mu)\phi$\\
Gamma&$-1/\mu$&$\mu^2$&$\mu^{-2}\mu'^2$&$(\phi-1)/(1+x)$\\
Inverse Gaussian&$-1/(2\mu^2)$&$\mu^3$&$\mu^{-3}\mu'^2$&$-\frac{3\mu^{3/2}}{2(\mu^{3/2}x+\mu)}+\frac{\phi\mu^{3/2}}{2(\mu^{3/2}x+\mu)^2}$\\
\hline
\end{tabular}
\label{tabela4}
\end{center}
\end{table}

One should notice that even though the support of the true residual of the inverse Gaussian distribution depends on the unknown parameter $\mu$, this fact does not have any implication with respect to the inferential aspect of the model. The regularity assumptions are satisfied since the estimation is done from the observed values of $Y_i$, $i=1,\ldots,n$. A discussion regarding the equality of the distribution of the corrected Pearson residual and the distribution of the true Pearson residual for the inverse Gaussian case is provided in Section 3 of Cordeiro and Simas (2009).

\section{Further adjustments on Pearson residuals}
In this section we recall the results obtained by Simas and Cordeiro (2009), and use their results to define a new adjusted residual, which we call PCA Pearson residuals. These residuals are inspired by the PCA residuals for normal linear models introduced by Rocha and Simas (2016). In the normal linear models they are independent and identically normally distributed, and can be used to construct exact Quantile-Quantile plots. The fact that the residuals are uncorrelated provide a sharper method to check for misspecification of the models. 
Therefore, we expect our new adjusted residual to have better asymptotic properties than the adjusted residuals defined by Simas and Cordeiro (2009).

These PCA Pearson residuals are new even for the generalized linear models. So we are introducing these residuals in a very general fashion.

\subsection{Moments of Pearson residuals}
Let $r = (r_1,\ldots,r_n)^T$ be the vector of $\mathcal{O}(n^{-1})$ expected values of the Pearson residuals. Simas and Cordeiro (2009)
used Cox and Snell's (1968) formulae to show that
\begin{equation}\label{expected}
r = -\frac{1}{2\phi} (I-H)(Jz+W^{1/2}d), 
\end{equation}
where $H=W^{1/2}\widetilde{X}(\widetilde{X}^TW\widetilde{X})^{-1}\widetilde{X}^TW^{1/2}$, $J = {\rm diag}\{V_i^{-1/2}\mu_i''\}$, $I$ is the identity
matrix of order $n$, $z = (z_{11},\ldots,z_{nn})^T$ is and $n\times 1$ vector with the diagonal elements of $Z$, and $d=(d_1,\ldots,d_n)^T$
is an $n\times 1$ vector with the diagonal elements of $D$.

Now, denote by $v = (v_1,\ldots,v_n)^T$ the vector of variances of the Pearson residuals to order $\mathcal{O}(n^{-1})$. Then,
Simas and Cordeiro (2009) showed that 
\begin{equation}\label{variance}
v = \frac{1}{\phi} \boldsymbol{1} + \frac{1}{2\phi^2} (QHJ-T)z + \frac{1}{2\phi^2}Q(H-I)W^{1/2}d,
\end{equation}
where $Q={\rm diag}\{V_i^{-1/2}V_i^{(1)}\}$ and $T={\rm diag}(2\phi w_i+ w_iV_i^{(2)}+V_i^{-1}V_i^{(1)}\mu_i''\}$.

Finally, let $i\neq j$, Simas and Cordeiro (2009) also showed that the covariance of order $\mathcal{O}(n^{-1})$ between $R_i$ and $R_j$
is given by
\begin{equation}\label{covariance}
{\rm Cov}(R_i,R_j) = -\frac{1}{\phi}h_{ij}, 
\end{equation}
where $h_{ij}$ is the $(i,j)$th element of $H$. Therefore, the covariance matrix of the Pearson residuals $R = (R_1,\ldots,R_n)^T$ is given
by $\Sigma = {\rm Cov}(R) = \{\sigma_{ij}\}$, where $\sigma_{ij} = \delta_{ij}v_i - \frac{1}{\phi}(1-\delta_{ij})h_{ij}$, and $\delta_{ij}$
is the Kronecker delta, which equals one if $i=j$ and zero if $i\neq j$.

Following Cordeiro (2004), Simas and Cordeiro (2009) defined adjusted Pearson residuals as
$$R_i^\ast = \frac{R_i - \hat{r}_i}{\hat{v}_i^{1/2}},$$
where $\hat{r}_i$ and $\hat{v}_i$ are obtained from the formulas for $r_i$ and $v_i$, respectively, by replacing $\mu_i$ by the fitted value
$\hat{\mu}_i$.
\subsection{The PCA Pearson residuals}
We now define new adjusted residuals. The point is that Cordeiro (2004) and Simas and Cordeiro (2009) argued that $R_i^\ast$ has better
normal approximation than $R_i$, because the convergence to the normal distribution is governed by the convergence of the first two moments. Nevertheless, they only used the asymptotic means and variances, so the second moments were used only marginally, that is, to correct the marginal distribution of the Pearson residuals. Our goal is to correct their joint distribution to yield a distribution closer to a multivariate normal distribution with a nearly diagonal covariance matrix. 

Thus, observe that the asymptotic covariance matrix of the vector of adjusted residuals 
$R^\ast = (R_1^\ast,\ldots,R_n^\ast)^T$ is given by the correlation matrix of the vector of Pearson residuals $R$,
and is given by
$$\Psi = {\rm diag}(v)^{-1/2}\, \Sigma\, {\rm diag}(v)^{-1/2}.$$

Following Rocha and Simas (2016), consider the spectral decomposition of $\Phi$: 
$$\Psi = E\Lambda E^{-1},$$
where $\Lambda = {\rm diag}\,\{\lambda_1,\lambda_2,\ldots,\lambda_n\}$, and $E$ is an orthogonal matrix of eigenvectors of $\Psi$.

Using this spectral decomposition we may define the PCA Pearson residuals as
\begin{equation}\label{pcares}
 \widetilde{R} = E^{-1} R^\ast,
\end{equation}
where $R^\ast = (R_1^\ast,\ldots,R_n^\ast)^T$ is the vector of adjusted Pearson residuals, and $E$ is the matrix in which each row
is an eigenvector of the correlation matrix $\Psi$. Observe that, since the matrix $E$ is orthogonal, we have $E^{-1} = E^T$.

We have directly that, to order $\mathcal{O}(n^{-1})$, $E(\widetilde{R}) = 0$ and $Cov(\widetilde{R}) = \Lambda$. First of all, it is expected that $\Psi$ is not of full rank. Let $n-m = rank(\Psi)$, thus the last $m$ eigenvalues of $\Psi$ are equal to zero. This means that, since $\widetilde{R}$ has approximately zero mean, to order $\mathcal{O}(n^{-1})$, $\widetilde{R}_{n-m+1} = \cdots = \widetilde{R}_n = 0$, where $\widetilde{R} = (\widetilde{R}_1,\ldots,\widetilde{R}_n)$. A well-specified model should have the remaining eigenvalues approximately equal, that is $\lambda_1\approx \cdots\approx \lambda_{n-m}$. Since our results are not exact, we follow the exact results by Rocha and Simas (2016) to suggest taking $m = p$, thus we study the first $n-p$ residuals, and disregard the remaining $p$ residuals.

Now, observe that the trace of the correlation matrix $\Psi$ will be approximately $n$, since all the adjusted residuals have, approximately, variance 1. Nevertheless, by the above reasoning, inspired by the results of Rocha and Simas (2016), we will have only $n-p$ nonzero residuals. Since the trace of a matrix is preserved under linear transformations, the result is that the sum of the variances of the $n-p$ nonzero PCA residuals will be $n$. Thus, each of the PCA Pearson residual will have, approximately, variance $n/(n-p)$, instead of 1. In order to avoid such discrepancy, we introduce the following modification of the PCA Pearson residuals:


\begin{equation}\label{pcaresmod}
 \breve{R} = \sqrt{\dfrac{n-p}{n}}\widetilde{R}.
\end{equation}

It is expected that both PCA residuals $\widetilde{R}$ and $\breve{R}$ have better normal approximation than the adjusted Pearson residuals $R_i^\ast$,
because not only $\widetilde{R}$ and $\breve{R}$ take into consideration the first two moments of $R_i$, it also takes into consideration their covariances, 
and thus, we are approximating the joint distribution of $\widetilde{R}$ and $\breve{R}$ by a multivariate normal. In this sense, the residuals
$\widetilde{R}$ and $\breve{R}$ use more information from the data, than the adjusted residuals $R^\ast$. 

The idea for the PCA residuals is the following: the eigenvector of the correlation matrix provides orthogonal directions with respect to
the covariance matrix, thus the PCA residuals will be orthogonal to order $\mathcal{O}(n^{-1})$, that is, they are approximately uncorrelated. If the assumption of approximate multivariate normal distribution holds, they will also be approximately independent.

Finally, since both residuals $\widetilde{R}$ and $\breve{R}$ are approximately uncorrelated, a quantile-quantile plot of these residuals against a normal
distribution should be more reliable than a Quantile-Quantile plot of the adjusted residuals $R^\ast$.
 
\section{Simulation results}
In this section our goal is to study the finite-sample distributions of
the Pearson residual $R_i$, its corrected version $R_i'$ obtained in Section 3, its adjusted version $R_i^\ast$ introduced by Simas and Cordeiro (2009),
and the PCA residuals $\widetilde{R}_i$ and $\breve{R}_i$, defined in Section 4. All simulations were performed using the statistical software \texttt{R}.

In this simulation experiment we consider a gamma nonlinear model with log link:
$$\log\mu_i = \beta_0 + x_{1,i}^{\beta_1}+\beta_2x_{2,i},\quad i=1,\ldots,n,$$
where the true parameters were taken as $\beta_0 = 1/2$, $\beta_1=1$, $\beta_2 = 2$ and $\phi=4$. The explanatory variables
$x_1$ and $x_2$ were generated from the uniform $U(0,1)$ distribution for $n=20$, and their values were held constant throughout
the simulations. The total number of Monte Carlo replications was set at $10,000$.

Note also that here the elements of the $n\times 3$ matrix $\widetilde{X}$ are $\widetilde{X}(\beta)_{i,1}=1,\widetilde{X}(\beta)_{i,2}=\log(x_{1,i})x_{1,i}^{\beta_1}$,
and $\widetilde{X}(\beta)_{i,3}=x_{2,i}$. The matrix $\widetilde{X}_i$ is given by
$$\widetilde{X}_i = \begin{bmatrix}
                 0&0&0\\
                 0&(\log(x_{1,i}))^2x_{1,i}^{\beta_1}&0\\
                 0&0&0
                \end{bmatrix}.$$
								
In each of the $10,000$ replications, we fitted the model and computed the MLE $\hat{\beta}$, fitted mean $\hat{\mu}$, the Pearson residual
$R_i$, $i=1,\ldots,15$, the functions $\rho_i(\cdot)$ using formula \eqref{formularho}, its corrected version $R_i'$, next we computed its expected value and covariance matrix from expressions \eqref{expected}, \eqref{variance} and \eqref{covariance},
and then the adjusted Pearson residual $R_i^\ast$ and the PCA Pearson residuals $\widetilde{R}_i$ and $\breve{R}_i$. 

In this case we have $e_i(x) = -1-x, h_i(x) = 1+x , w=1$, thus
$$\rho_i(x) = (1+x)\left\{ \gamma_i^T \widetilde{X} B(\hat{\beta}) + \frac{d_i}{2\phi} + \frac{z_{ii}}{2}x\right\}.$$

Table \ref{tabelamoments} gives the sample mean, variance, skewness and kurtosis of the residuals $R_i, R_i'$ and $\varepsilon_i$ out of
$10,000$ values, and in this case we may use the figures for $\varepsilon_i$ as benchmark. The theoretical values for the mean, variance, skewness and kurtosis for
the distribution of the true Pearson residuals for $\phi=4$ (for the density of such distribution see Table \ref{tabela1}) are, respectively, 0, 0.25, 1 and 4.5.

\begin{center}
\begin{table}[htbp]
\caption{Moments of the uncorrected ($R_i$), corrected ($R_i'$) and true $(\varepsilon_i)$ Pearson residuals.}
\centering
\scriptsize
\begin{tabular}{|c|rrr|rrr|rrr|rrr|}
\hline
\multirow{ 2}{*}{$i$} & \multicolumn{3}{c|}{Mean}&\multicolumn{3}{c|}{Variance}& \multicolumn{3}{c|}{Skewness}&\multicolumn{3}{c|}{Kurtosis} \\ 
\cline{2-13}
& $R_i$ & $R_i'$ & $\varepsilon_i$ &$R_i$ & $R_i'$ & $\varepsilon_i$& $R_i$ & $R_i'$ & $\varepsilon_i$ &$R_i$ & $R_i'$ & $\varepsilon_i$\\ \hline
1 & $-0.065 $ &	$-0.025$   & $-0.001$ &  $0.189$   & $0.225$ &  $0.249$	  &	$ 0.747$	 & $0.904$   &	$1.007$  &  $3.746$	      &	 $4.291$   &  $4.699$ \\   
2 & $ 0.074 $ &	$ 0.013$   & $ 0.007$ &  $0.258$   & $0.246$ &  $0.256$	  &	$ 0.836$	 & $0.949$   &	$1.009$  &  $3.953$ 	  &	 $4.322$   &  $4.550$ \\	
3 & $ 0.023 $ &	$ 0.004$   & $ 0.001$ &  $0.211$   & $0.238$ &  $0.254$	  &	$ 0.718$	 & $0.913$   &	$1.031$  &  $3.660$ 	  &	 $4.274$   &  $4.714$ \\	
4 & $-0.055 $ &	$-0.024$   & $-0.003$ &  $0.189$   & $0.228$ &  $0.249$	  &	$ 0.728$	 & $0.906$   &	$0.998$  &  $3.647$ 	  &	 $4.212$   &  $4.529$ \\   
5 & $ 0.038 $ &	$-0.002$   & $ 0.004$ &  $0.212$   & $0.230$ &  $0.253$	  &	$ 0.710$	 & $0.871$   &	$1.041$  &  $3.613$ 	  &	 $3.942$   &  $4.800$ \\   
6 & $ 0.029 $ &	$-0.010$   & $-0.001$ &  $0.209$   & $0.223$ &  $0.249$	  &	$ 0.676$	 & $0.815$   &	$0.980$  &  $3.497$ 	  &	 $3.747$   &  $4.461$ \\   
7 & $-0.048 $ &	$-0.010$   & $ 0.012$ &  $0.195$   & $0.239$ &  $0.261$	  &	$ 0.751$	 & $0.914$   &	$1.017$  &  $3.636$ 	  &	 $4.124$   &  $4.577$ \\   
8 & $ 0.064 $ &	$-0.000$   & $ 0.000$ &  $0.240$   & $0.234$ &  $0.250$	  &	$ 0.798$	 & $0.927$   &	$1.037$  &  $3.912$ 	  &	 $4.272$   &  $4.811$ \\   
9 & $-0.021 $ &	$-0.015$   & $-0.006$ &  $0.179$   & $0.226$ &  $0.252$	  &	$ 0.615$	 & $0.873$   &	$0.993$  &  $3.371$ 	  &	 $4.136$   &  $4.463$ \\   
10& $ 0.042 $ &	$ 0.002$   & $-0.003$ &  $0.215$   & $0.229$ &  $0.246$	  &	$ 0.688$	 & $0.852$   &	$0.977$  &  $3.517$ 	  &	 $3.972$   &  $4.344$ \\   
11& $ 0.044 $ &	$ 0.005$   & $ 0.000$ &  $0.196$   & $0.222$ &  $0.242$	  &	$ 0.592$	 & $0.806$   &	$0.923$  &  $3.278$ 	  &	 $3.784$   &  $4.204$ \\   
12& $-0.036 $ &	$-0.016$   & $-0.005$ &  $0.195$   & $0.228$ &  $0.243$	  &	$ 0.739$	 & $0.895$   &	$0.987$  &  $3.702$ 	  &	 $4.227$   &  $4.518$ \\   
13& $ 0.046 $ &	$ 0.003$   & $-0.000$ &  $0.226$   & $0.240$ &  $0.248$	  &	$ 0.740$	 & $0.929$   &	$0.983$  &  $3.654$ 	  &	 $4.216$   &  $4.460$ \\   
14& $-0.082 $ &	$-0.035$   & $-0.001$ &  $0.169$   & $0.217$ &  $0.251$	  &	$ 0.629$	 & $0.848$   &	$0.980$  &  $3.322$ 	  &	 $3.905$   &  $4.371$ \\   
15& $-0.004 $ &	$-0.010$   & $-0.010$ &  $0.223$   & $0.238$ &  $0.248$	  &	$ 0.826$	 & $0.922$   &	$1.022$  &  $3.822$ 	  &	 $4.127$   &  $4.572$ \\   
16& $ 0.055 $ &	$ 0.001$   & $ 0.001$ &  $0.237$   & $0.236$ &  $0.247$	  &	$ 0.751$	 & $0.864$   &	$0.979$  &  $3.579$ 	  &	 $3.863$   &  $4.363$ \\   
17& $-0.001 $ &	$-0.010$   & $ 0.003$ &  $0.170$   & $0.211$ &  $0.239$	  &	$ 0.533$	 & $0.778$   &	$0.897$  &  $3.130$ 	  &	 $3.700$   &  $4.102$ \\   
18& $-0.029 $ &	$-0.005$   & $ 0.006$ &  $0.209$   & $0.239$ &  $0.254$	  &	$ 0.801$	 & $0.920$   &	$0.984$  &  $3.685$ 	  &	 $4.108$   &  $4.251$ \\   
19& $ 0.005 $ &	$-0.000$   & $-0.001$ &  $0.219$   & $0.240$ &  $0.252$	  &	$ 0.836$	 & $0.960$   &	$1.052$  &  $3.812$ 	  &	 $4.200$   &  $4.628$ \\   
20& $-0.079 $ &	$-0.040$   & $-0.008$ &  $0.173$   & $0.208$ &  $0.245$	  &	$ 0.642$	 & $0.803$   &	$0.978$  &  $3.344$ 	  &	 $3.738$   &  $4.468$ \\ 
\hline                                                                                             
\end{tabular}
\label{tabelamoments}
\end{table}
\end{center}

The figures in Table \ref{tabelamoments} show that the distribution of the corrected Pearson residuals is generally closer to the
distribution of the true residuals than the distribution of the Pearson residuals. The correction function $\rho(\cdot)$ seems to be effective
even when the sample size is small. The distribution of all residuals for the gamma model are positively skewed, as indicated from the theoretical skewness, which is 1. All four cumulants of the corrected Pearson residuals $R_i'$ are generally closer to the corresponding cumulants of the true Pearson residuals $\varepsilon_i$ than those of the Pearson residuals.

Table \ref{tabelaksone} gives the values and respective $p$-values of the one-sample Kolmogorov-Smirnov (K-S) distance between the empirical distribution of the
uncorrected and corrected residuals and the estimated distribution of the true residuals (a shifted gamma). The values of the
K-S statistic measure the distances between the estimated distribution of the true residuals $\varepsilon_i$ and the empirical distribution of each set of 10,000 uncorrected Pearson residuals
$R_i$ and corrected Pearson residuals $R_i'$, for $i = 1 ,\ldots, 20$. Here, the estimated
distribution is the shifted gamma distribution with dispersion parameter $\phi$ estimated by the sample mean of the estimates
of the dispersion parameter at each step of the Monte Carlo experiment.

We are also interested in checking whether the empirical distributions of the uncorrected 
$R_i$ and corrected $R_i'$ residuals agrees with the empirical distribution of the true residuals $\varepsilon_i$. 
Hence, we give in Table \ref{tabelakstwo} values of two-sample K-S statistic, with their respective $p$-values, 
between the empirical distribution of the uncorrected and corrected Pearson residuals and the empirical distribution of the true
residuals.

\begin{center}
\begin{table}[htbp]
\caption{One-sample K-S statistics and $p$-values on Pearson and corrected Pearson residuals.}
\centering
\scriptsize
\begin{tabular}{|c|ll|ll|}
\hline
\multirow{ 2}{*}{$i$} & \multicolumn{2}{c|}{K-S Statistic}&\multicolumn{2}{c|}{K-S $p$-values} \\ 
\cline{2-5}
& $R_i$ & ${R}_i'$ & ${R}_i$ &$R_i'$\\ \hline
1 &$0.038$ & $0.012$&$ 1.479\times 10^{-13} $ & $0.093                $\\  					   
2 &$0.064$ & $0.012$&$ 0.000                $ & $0.092                $\\     				   
3 &$0.025$ & $0.010$&$ 7.066\times 10^{-6}  $ & $0.226              	 $\\ 				   
4 &$0.061$ & $0.015$&$ 0.000                $ & $0.013                $\\					   
5 &$0.073$ & $0.012$&$ 0.000                $ & $0.093                $\\   				   
6 &$0.043$ & $0.013$&$ 1.110\times 10^{-16} $ & $0.067                $\\      				   	
7 &$0.034$ & $0.008$&$ 1.553\times 10^{-10} $ & $0.476                $ \\       			   
8 &$0.051$ & $0.021$&$ 0.000                $ & $1.911\times 10^{-4}  $\\                       
9 &$0.068$ & $0.027$&$ 0.000                $ & $4.468\times 10^{-7}  $\\                       
10&$0.052$ & $0.020$&$ 0.000                $ & $5.077\times 10^{-7}  $\\                       
11&$0.037$ & $0.010$&$ 2.306\times 10^{-12} $ & $0.252                $\\      				   
12&$0.055$ & $0.017$&$ 0.000                $ & $0.005                $\\      				   
13&$0.045$ & $0.010$&$ 0.000                $ & $0.227                $\\         			   
14&$0.060$ & $0.017$&$ 0.000                $ & $0.005                $\\     				   
15&$0.060$ & $0.024$&$ 0.000                $ & $1.464\times 10^{-5}  $\\                      
16&$0.040$ & $0.009$&$ 6.106\times 10^{-15} $ & $0.309                $\\      				   
17&$0.047$ & $0.006$&$ 0.000                $ & $0.756                $\\          			   
18&$0.019$ & $0.009$&$ 9.379\times 10^{-4}  $ & $0.369                $\\       			   
19&$0.066$ & $0.024$&$ 0.000                $ & $8.580\times 10^{-6}  $\\                      
20&$0.030$ & $0.012$&$ 1.603\times 10^{-8}  $ & $0.089 				 $\\  
\hline                                                                                             
\end{tabular}
\label{tabelaksone}
\end{table}
\end{center}

\begin{center}
\begin{table}[htbp]
\caption{Two-sample K-S statistics and $p$-values on Pearson and corrected Pearson residuals.}
\centering
\scriptsize
\begin{tabular}{|c|ll|ll|}
\hline
\multirow{ 2}{*}{$i$} & \multicolumn{2}{c|}{K-S Statistic}&\multicolumn{2}{c|}{K-S $p$-values} \\ 
\cline{2-5}
& $R_i$ & ${R}_i'$ & ${R}_i$ &$R_i'$\\ \hline
1 & $0.051$& $7.442\times 10^{-12} $&$0.020$&$ 0.025$                    \\   
2 & $0.071$& $0.000                $&$0.015$&$ 0.198$                    \\	
3 & $0.045$& $2.041 \times 10^{-9} $&$0.014$&$ 0.216$                    \\	
4 & $0.043$& $1.570 \times 10^{-8} $&$0.017$&$ 0.111$                    \\   
5 & $0.054$& $1.815 \times 10^{-13}$&$0.012$&$ 0.395$                    \\   
6 & $0.057$& $1.232 \times 10^{-14}$&$0.012$&$ 0.405$     \\   
7 & $0.048$& $1.971 \times 10^{-10}$&$0.019$&$ 0.052$       \\   
8 & $0.072$& $0.000                $&$0.011$&$ 0.568$     \\   
9 & $0.039$& $3.094 \times 10^{-7} $&$0.016$&$ 0.144$   \\   
10& $0.065$& $0.000                $&$0.017$&$ 0.084$        \\   
11& $0.072$& $0.000                $&$0.020$&$ 0.025$      \\   
12& $0.032$& $4.541 \times 10^{-5} $&$0.013$&$ 0.348$       \\   
13& $0.063$& $0.000                $&$0.013$&$ 0.339$        \\   
14& $0.071$& $0.000                $&$0.029$&$ 3.525 \times 10^{-5}$   \\   
15& $0.026$& $0.001          	  $	&$0.011$&$ 0.534$      \\
16& $0.061$& $1.110 \times 10^{-16}$&$0.008$&$ 0.889$      \\   
17& $0.048$& $1.626 \times 10^{-10}$&$0.016$&$ 0.149$           \\   
18& $0.029$& $4.201 \times 10^{-4} $&$0.013$&$ 0.366$             \\   
19& $0.032$& $4.541 \times 10^{-5} $&$0.012$&$ 0.456$              \\   
20& $0.065$& $0.000                $&$0.029$&$ 3.525 \times 10^{-5}$ \\      
\hline                                                                                             
\end{tabular}
\label{tabelakstwo}
\end{table}
\end{center}

The figures in Tables \ref{tabelaksone} and \ref{tabelakstwo} indicate that the empirical distributions of the corrected residuals $R_i'$ are much closer to the
distribution of the true residuals than the empirical distributions of the uncorrected residuals $R_i$. Indeed, for both the one-sample and two-sample K-S tests, all the $p$-values regarding the uncorrected Pearson residuals $R_i$ were smaller than the usual significance level of $5\%$. Thus indicating that the uncorrected Pearson residuals do not follow the same distribution as the true residuals. By looking at the $p$-values for the corrected Pearson residuals, we observe that for the one-sample K-S test, we obtained, at the usual significance level of $5\%$, 12 out of 20 residuals following the shifted gamma distribution (the theoretical distribution of the true residuals), and for the two-sample K-S test, we obtained, at the usual significance level of $5\%$, 16 out of 20 residuals following the same distribution as the true residual (their empirical distribution). This fact indicates that the corrected residuals represent a considerable improvement over the uncorrected residuals when the model is well-specified. Indeed, one should observe that the 8 out of 20 residuals that we assumed not to follow the shifted gamma distribution, based on the one-sample K-S test, had much smaller distances to such distribution when compared to the uncorrected residuals. The same phenomenon occurs when we look at the 4 out of 20 residuals that we assumed not to follow the same distribution as the true residuals, based on the two-sample K-S test.

Furthermore, we will now compute, out of the total of 10,000 simulated data sets, the proportion of the Kolmogorov-Smirnov tests that did reject the null hypothesis that the residuals follow the shifted-gamma distribution, based on each simulated dataset of 20 observations. More precisely, for each $i$, $i$ ranging from 1 to 10,000, we will obtain the $p$-value of the Kolmogorov-Smirnov test of the distribution of the set of 20 residuals against the distribution of the true residual (in this case, a shifted gamma distribution). Then we will compare each $p$-value to the following significance levels: 1\%, 2.5\%, 5\%, 7.5\%, 10\%, 12.5\% and 15\%. Afterwards, we compute the proportion of the rejected tests for each significance level and compare these proportions to their corresponding nominal level. The results are provided in Table \ref{tabelaksdataset1} below:

\begin{center}
\begin{table}[htbp]
\caption{Proportions of rejected K-S tests based on each simulated dataset for different significance levels.}
\centering
\scriptsize
\begin{tabular}{c|c|c|c|c|c|c|c}
\hline
Residual & 1\% level & 2.5\% level & 5\% level & 7.5\% level & 10\% level & 12.5\% level & 15\% level\\
\hline
$R$ & $2\times 10^{-4}$ & $3\times 10^{-4}$&$9\times 10^{-4}$&0.0033&0.0072&0.012&0.0173\\
$R'$ & 0 & $2\times 10^{-4}$&$6\times 10^{-4}$&0.0026&0.0051&0.0083&0.0133\\
\hline                                                                                             
\end{tabular}
\label{tabelaksdataset1} 
\end{table}
\end{center}

By looking at the results presented in Table \ref{tabelaksdataset1} above, we conclude that, for all residuals, the empirical significance levels are very far from the nominal levels. This result is expected for the uncorrected Pearson residual, since, as the previous results have shown, their marginal distributions are not shifted gamma distributions. Thus, one should not expect them to form a random sample of a shifted gamma distribution. The surprise comes when we observe that the same phenomenon occurs with the corrected residuals. We do not obtain the corresponding nominal level, indicating that the set of corrected residuals also do not form a random sample of shifted gamma distributions. But, by looking closely we see that this comes from the fact that these residuals are not independent, so the dependence among them, cause their joint distribution to be different from an independent and identically distributed (iid) shifted gamma sample. The above results show that the K-S test based on both the uncorrected and correct Pearson residual tend to be too optimistic. 

We now move to the analysis of the PCA Pearson residuals $\widetilde{R}_i$ and $\breve{R}_i$ along with the adjusted Pearson residual introduced by Simas and Cordeiro (2009).

Table \ref{tabelamoments2} gives the sample mean, variance, skewness and kurtosis of the residuals $R_i^\ast, \widetilde{R}_i$ and $\breve{R}_i$ out of
$10,000$ values. For these residuals we are looking for a good agreement with the normal distribution. This happens when those figures are close to $0,1,0$ and $3$, respectively. 

We begin by recalling the remark made in Section 4 where we argued that one should consider the first $n-p = 17$ PCA Pearson residuals and disregard the last $p=3$ residuals. 

\begin{center}
\begin{table}[htb!]
\caption{Moments of the adjusted Pearson residual $(R_i^\ast)$, PCA Pearson residual $\widetilde{R}_i$ and PCA Pearson residual $\breve{R}_i$.}
\centering
\scriptsize
\begin{tabular}{|c|rrr|rrr|rrr|rrr|}
\hline
\multirow{ 2}{*}{$i$} & \multicolumn{3}{c|}{Mean}&\multicolumn{3}{c|}{Variance}& \multicolumn{3}{c|}{Skewness}&\multicolumn{3}{c|}{Kurtosis} \\ 
\cline{2-13}
& $R_i^\ast$ & $\widetilde{R}_i$ & $\breve{R}_i$ &$R_i^\ast$ & $\widetilde{R}_i$ & $\breve{R}_i$& $R_i^\ast$ & $\widetilde{R}_i$ & $\breve{R}_i$ &$R_i^\ast$ & $\widetilde{R}_i$ & $\breve{R}_i$\\ \hline
1 & $-0.030 $ &	$ -0.008$   & $-0.007$ &  $0.986$   & $1.445$ &  $1.228$	  & $0.731$	 & $ 0.230$   &	$ 0.230$  &  $ 3.677$     &	 $ 3.140$   &  $3.140$ \\   
2 & $ 0.037 $ &	$  0.003$   & $ 0.002$ &  $0.999$   & $1.339$ &  $1.138$	  & $0.798$	 & $ 0.045$   &	$ 0.045$  &  $ 3.842$ 	  &	 $ 3.159$   &  $3.159$ \\	
3 & $ 0.016 $ &	$  0.003$   & $ 0.003$ &  $0.999$   & $1.249$ &  $1.061$	  & $0.693$	 & $ 0.008$   &	$ 0.008$  &  $ 3.577$ 	  &	 $ 2.957$   &  $2.957$ \\	
4 & $-0.031 $ &	$ -0.001$   & $-0.001$ &  $1.000$   & $1.250$ &  $1.062$	  & $0.736$	 & $ 0.006$   &	$ 0.006$  &  $ 3.707$ 	  &	 $ 3.145$   &  $3.145$ \\   
5 & $ 0.009 $ &	$ -0.001$   & $-0.001$ &  $0.984$   & $1.185$ &  $1.007$	  & $0.650$	 & $ 0.024$   &	$ 0.024$  &  $ 3.372$ 	  &	 $ 3.038$   &  $3.038$ \\   
6 & $-0.009 $ &	$  0.015$   & $ 0.014$ &  $0.951$   & $1.159$ &  $0.985$	  & $0.626$	 & $ 0.034$   &	$ 0.034$  &  $ 3.291$ 	  &	 $ 3.116$   &  $3.116$ \\   
7 & $ 0.001 $ &	$  0.010$   & $ 0.010$ &  $1.058$   & $1.174$ &  $0.998$	  & $0.728$	 & $-0.023$   &	$-0.023$  &  $ 3.566$ 	  &	 $ 3.178$   &  $3.178$ \\   
8 & $ 0.010 $ &	$  0.010$   & $ 0.009$ &  $0.961$   & $1.154$ &  $0.981$	  & $0.749$	 & $-0.025$   &	$-0.025$  &  $ 3.706$ 	  &	 $ 3.154$   &  $3.154$ \\   
9 & $-0.009 $ &	$ -0.012$   & $-0.011$ &  $1.052$   & $1.183$ &  $1.006$	  & $0.607$	 & $ 0.008$   &	$ 0.008$  &  $ 3.374$ 	  &	 $ 3.190$   &  $3.190$ \\   
10& $ 0.017 $ &	$ -0.018$   & $-0.016$ &  $0.960$   & $1.102$ &  $0.937$	  & $0.643$	 & $-0.021$   &	$-0.021$  &  $ 3.420$ 	  &	 $ 3.304$   &  $3.304$ \\   
11& $ 0.031 $ &	$ -0.013$   & $-0.012$ &  $0.981$   & $1.129$ &  $0.960$	  & $0.556$	 & $-0.004$   &	$-0.004$  &  $ 3.189$ 	  &	 $ 3.151$   &  $3.151$ \\   
12& $-0.020 $ &	$ -0.005$   & $-0.005$ &  $0.976$   & $1.091$ &  $0.927$	  & $0.725$	 & $-0.010$   &	$-0.010$  &  $ 3.677$ 	  &	 $ 3.229$   &  $3.229$ \\   
13& $ 0.013 $ &	$  0.000$   & $ 0.000$ &  $0.986$   & $1.127$ &  $0.958$	  & $0.723$	 & $-0.065$   &	$-0.065$  &  $ 3.622$ 	  &	 $ 3.413$   &  $3.413$ \\   
14& $-0.029 $ &	$ -0.001$   & $-0.001$ &  $1.093$   & $1.100$ &  $0.935$	  & $0.631$	 & $ 0.112$   &	$ 0.112$  &  $ 3.331$ 	  &	 $ 3.269$   &  $3.269$ \\   
15& $-0.014 $ &	$  0.016$   & $ 0.015$ &  $0.977$   & $1.095$ &  $0.931$	  & $0.803$	 & $ 0.130$   &	$ 0.130$  &  $ 3.751$ 	  &	 $ 3.505$   &  $3.505$ \\   
16& $ 0.011 $ &	$  0.008$   & $ 0.008$ &  $0.968$   & $1.049$ &  $0.892$	  & $0.708$	 & $ 0.066$   &	$ 0.066$  &  $ 3.439$ 	  &	 $ 3.589$   &  $3.589$ \\   
17& $ 0.004 $ &	$  0.000$   & $ 0.000$ &  $1.018$   & $1.036$ &  $0.880$	  & $0.512$	 & $ 0.131$   &	$ 0.131$  &  $ 3.163$ 	  &	 $ 3.890$   &  $3.890$ \\   
18& $-0.000 $ &	$  0.006$   & $ 0.006$ &  $1.005$   & $0.011$ &  $0.010$	  & $0.794$	 & $ 0.025$   &	$ 0.025$  &  $ 3.731$ 	  &	 $ 4.434$   &  $4.434$ \\   
19& $ 0.005 $ &	$  0.003$   & $ 0.003$ &  $0.989$   & $0.016$ &  $0.014$	  & $0.809$	 & $ 0.161$   &	$ 0.161$  &  $ 3.737$ 	  &	 $ 4.146$   &  $4.146$ \\   
20& $-0.051 $ &	$  0.000$   & $ 0.000$ &  $0.973$   & $0.021$ &  $0.018$	  & $0.629$	 & $ 0.042$   &	$ 0.042$  &  $ 3.304$ 	  &	 $ 4.907$   &  $4.907$ \\    
\hline                                                                                             
\end{tabular}
\label{tabelamoments2}
\end{table}
\end{center}

It is noteworthy that the results presented in Table \ref{tabelamoments2} corroborate that remark. Indeed, the last 3 residuals had both mean and variance close to 0, thus suggesting an approximate constant value of zero. 

The figures in Table \ref{tabelamoments2} show us that: the mean of all residuals are close to zero; the variance of the adjusted Pearson residuals are generally closer to 1 than those of the PCA Pearson residuals (here, we disregard the last 3 residuals); the adjusted Pearson residuals are highly positively skewed, thus indicating a departure of normality, whereas the PCA Pearson residuals are approximately unskewed; the adjusted Pearson residuals have a high excess kurtosis, more precisely, 14 out of 20 (70\%) adjusted Pearson residuals had kurtosis higher than 3.4, whereas for both PCA Pearson residuals the number of residuals with kurtosis higher than 3.4 were 4 out of 17 (23.53\%). 

These figures thus suggest that the adjusted Pearson residual seem to provide a better correction of the first two moments, nevertheless, its overall normality assumption seems to be inadequate. With respect to the PCA Pearson residuals, the overall normality assumption seems adequate, but as expected from its Principal Component Analysis nature, we had a ranking of variances, and, even though those variances were supposed to be equal, in practice they are slightly different, thus affecting the variances of the different residuals. Since we are interested in normality approximation instead of simply correcting the first two moments, we observe that the results in Table \ref{tabelamoments2} suggest that both PCA residuals $\widetilde{R}_i$ and $\breve{R}_i$ should provide a better normal approximation than the adjusted Pearson residuals.  

We will now verify that claim that both PCA Pearson residuals provide better normal approximation than the adjusted Pearson residuals by means of the one-sample Kolmogorov-Smirnov test against a standard normal distribution. 

Table \ref{tabelaksnorm}  gives the statistics and $p$-values of the one-sample Kolmogorov-Smirnov test for $R_i^\ast$, $\widetilde{R}_i$ and $\breve{R}_i$, for each $i=1,\ldots,20$, against the theoretical quantiles of a standard normal distribution, namely $N(0,1)$. The values of the
K-S statistic measure the distances between the standard normal distribution and the empirical distributions of each set of 10,000 adjusted Pearson residuals
$R_i^\ast$, PCA Pearson residuals $\widetilde{R}_i$, and PCA Pearson residuals $\breve{R}_i$, for $i = 1 ,\ldots, 20$. 

%

\begin{center}
\begin{table}[htbp]
\caption{One-sample K-S statistics and $p$-values on adjusted and both PCA Pearson residuals.}
\centering
\scriptsize
\begin{tabular}{|c|lll|lll|}
\hline
\multirow{ 2}{*}{$i$} & \multicolumn{3}{c|}{K-S Statistic}&\multicolumn{3}{c|}{K-S $p$-values} \\ 
\cline{2-7}
& $R_i^\ast$ & $\widetilde{R}_i$ & $\breve{R}_i$ &$R_i^\ast$ & $\widetilde{R}_i$ & $\breve{R}_i$\\ \hline
1 & $0.063 $& $0.056$  &$	0.037$  & $ 0.000     	        $            & $0.000$                 & $7.142\times 10^{-13}$          		  \\ 								 
2 & $0.042 $& $0.038$  &$ 0.014$ & $ 2.220\times 10^{-16}  $          & $1.458\times 10^{-13}$   & $0.045 $        \\	   									 
3 & $0.043 $& $0.033$  &$	0.013$  & $ 1.110\times 10^{-16}  $          & $4.776\times 10^{-10}$  & $0.054         		$		  \\										 
4 & $0.062 $& $0.028$  &$	0.011$  & $ 0.000     	        $            & $2.624\times 10^{-7}$   & $0.160         		$		  \\ 										 
5 & $0.043 $& $0.026$  &$	0.008$  & $ 1.110\times 10^{-16}  $          & $2.198\times 10^{-6}$   & $0.490       		$		  \\ 										 
6 & $0.050 $& $0.021$  &$	0.010$  & $ 0.000       	        $        & $1.291\times 10^{-4}$   & $0.269         		$		  \\ 										 
7 & $0.053 $& $0.020$  &$	0.009$  & $ 0.000      	        $            & $4.527\times 10^{-4}$   & $0.298          		$	  \\ 											 
8 & $0.049 $& $0.022$  &$	0.008$  & $ 0.000      	        $            & $1.074\times 10^{-4}$   & $0.426        		$		  \\ 										 
9 & $0.050 $& $0.021$  &$	0.013$  & $ 0.000      	        $            & $2.621\times 10^{-4}$   & $0.059          		$	  \\ 											 
10& $0.039 $& $0.012$  &$	0.012$  & $ 5.240\times 10^{-14}  $          & $0.068$                 & $0.072 $     				  \\ 							 
11& $0.031 $& $0.017$  &$	0.013$  & $ 3.695\times 10^{-9}   $          & $0.005$                 & $0.053        		$					  \\ 							 
12& $0.056 $& $0.011$  &$	0.012$  & $ 0.000   	            $        & $0.125$                 & $0.069       		$					  \\ 							 
13& $0.045 $& $0.013$  &$	0.012$  & $ 0.000   	            $        & $0.045$                 & $0.075         		$					  \\ 							 
14& $0.059 $& $0.013$  &$	0.016$  & $ 0.000   	            $        & $0.040$                 &  $0.006            	$					  \\ 							 
15& $0.069 $& $0.012$  &$	0.021$  & $ 0.000   	            $        & $0.098$                 & $1.636\times 10^{-4} $         			  \\ 							 
16& $0.048 $& $0.007$  &$	0.025$  & $ 0.000   	            $        & $0.637$                 & $3.981\times 10^{-6} $            		  \\ 							 
17& $0.038 $& $0.008$  &$	0.026$  & $ 2.664\times 10^{-13}  $          & $0.399$                 & $1.167\times 10^{-6} $       			  \\ 							 
18& $0.060 $& $0.387$  &$	0.393$  & $ 0.000    	            $        & $0.000$                 & $0.000          $					  \\ 								 
19& $0.061 $& $0.375$  &$	0.383$  & $ 0.000    	            $        & $0.000$                 &  $0.000         	$				  \\ 									 
20& $0.068 $& $0.356$  &$	0.365$  & $ 0.000    	            $        & $0.000$                 & $0.000			$					  \\ 				
\hline                                                                                             
\end{tabular}
\label{tabelaksnorm} 
\end{table}
\end{center}

The figures in Table \ref{tabelaksnorm} indicate that the empirical distribution of the PCA Pearson residuals $\breve{R}_i$ are much closer to the
standard normal distribution than the empirical distribution of both adjusted Pearson residuals $R_i^\ast$ and PCA Pearson residuals $\widetilde{R}_i$. Indeed, all the $p$-values regarding the adjusted Pearson residuals $R_i^\ast$ were smaller than the usual significance level of $5\%$. Thus indicating that the adjusted Pearson residuals do not follow the standard normal distribution. By looking at the $p$-values for the PCA Pearson residuals $\breve{R}_i$, we observe that we obtained, at the usual significance level of $5\%$, 11 out of 17 residuals following the standard normal distribution, and for the PCA Pearson residuals $\widetilde{R}_i$, we obtained, at the usual significance level of $5\%$, 5 out of 17 residuals following the standard normal distribution. This fact indicates that both PCA Pearson residuals represent a considerable improvement over the adjusted Pearson residuals in terms of normal approximation when the model is well-specified. Indeed, one should observe that for each case of both PCA Pearson residuals, the K-S statistic were smaller than those of the adjusted Pearson residual, thus indicating that both PCA Pearson residuals are closer to the standard normal distribution than the adjusted Pearson residual. 

Therefore, our general conclusion up to this point is that the PCA Pearson residual $\breve{R}_i$ provides the best normal approximation among the studied residuals. Thus, we suggest the usage of such residual to check model assumptions.

Finally, we will now compute, out of the total of 10,000 simulated data sets, the proportion of the Kolmogorov-Smirnov tests that did reject the null hypothesis that the residuals follow the standard normal distribution, based on each simulated dataset of 20 observations for the adjusted Pearson residuals, and 17 observations for the PCA Pearson residuals. More precisely, for each $i$, $i$ ranging from 1 to 10,000, we will obtain the $p$-value of the Kolmogorov-Smirnov test of the distribution of the set of 20 residuals (and 17 residuals for the PCA residuals) against the standard normal distribution. Then we will compare each $p$-value to the following significance levels: 1\%, 2.5\%, 5\%, 7.5\%, 10\%, 12.5\% and 15\%. Afterwards, we compute the proportion of the rejected tests for each significance level and compare these proportions to their corresponding nominal level. The results are provided in Table \ref{tabelaksdataset2} below:


%

%

\begin{center}
\begin{table}[htbp]
\caption{KS statistics and $p$-values}
\centering
\scriptsize
\begin{tabular}{c|c|c|c|c|c|c|c}
\hline
Residual & 1\% level & 2.5\% level & 5\% level & 7.5\% level & 10\% level & 12.5\% level & 15\% level\\
\hline
$R_i^\ast$        &$2\times 10^{-4}$ &0.001 &0.0039 &0.009&0.0152 &0.0227 &0.0312 \\
$\widetilde{R}_i$ & 0.0123&0.0302 &0.0552 &0.0824 &0.1065 &0.1317 &0.1583 \\
$\breve{R}_i$     &0.0099 &0.0246 &0.0461 &0.0705 &0.0942 &0.1191 &0.1428 \\
\hline                                                                                             
\end{tabular}
\label{tabelaksdataset2} 
\end{table}
\end{center}

By looking at the results presented in Table \ref{tabelaksdataset2} above, we observe that the empirical significance level of the adjusted Pearson residuals are very far from the nominal level. This is an expected result since, as the previous results have shown, their marginal distributions are standard normal. Thus, one should not expect them to form a random sample of standard normal distribution. Nevertheless, when we move to the empirical significance levels of the PCA Pearson residuals, we see that they are very close to the nominal levels, with the PCA Pearson residuals $\breve{R}_i$ having the best results. Indeed, for each significance level we considered, the empirical significance levels of the PCA Pearson residuals $\breve{R}_i$ were closer to the nominal level than the PCA Pearson residual $\widetilde{R}_i$. Thus, this indicates that the PCA Pearson residuals $\breve{R}_i$ do form, jointly, a random sample of standard normal distribution. This comes from the fact that the PCA Pearson residuals are approximately uncorrelated, and since they are approximately normally distributed, as seen in the previous results, they are approximately independent. This shows that not only the Kolmogorov-Smirnov tests using these PCA Pearson residuals are more reliable than the remaining residuals considered here, but also the Quantile-Quantile plots should also be more reliable, since they do form, approximately, a random rample of a standard normal distribution.




%
%
%

\section{Conclusions}
In exponential family regression models, Pearson residuals are either compared with quantiles of the standard normal distribution or
analyzed with the aid of residual plots with simulated envelopes. However, the normal approximation is not adequate
in small samples, even for the linear case, as seen in Cordeiro and Simas (2009). To circumvent this issue we tackled the problem from two different directions.
At first, we defined corrected residuals for these models which have the same distribution of the true residuals to order $\mathcal{O}(n^{-1})$. The setup is similar to the paper by Loynes (1969), and extends the previous result by Cordeiro and Simas (2009), and thus can also be considered as a sequel to such article. We provide tables to aid applications to some common models. The performance of the uncorrected and the corrected
Pearson residuals are compared in a simulation study under a well-specified gamma model. The simulation results show that, as expected, the corrected Pearson residuals can be assumed to follow the same distribution as the true Pearson residuals, whereas the uncorrected residuals do not follow such distribution. 
Secondly, we defined two PCA Pearson residuals, namely, $\widetilde{R}$ and $\breve{R}$. Their performance with respect to standard normal approximation were
compared to that of the adjusted Pearson residuals introduced by Simas and Cordeiro (2009) by simulation under a well-specified gamma model. The simulation results showed that the PCA Pearson residual $\breve{R}$ had the best performance. Indeed, as the simulation results show the PCA Pearson residual $\breve{R}$ can be assumed to follow a standard normal distribution, whereas the adjusted Pearson residuals, as can be seen in Simas and Cordeiro (2009), can be seen as an improvement of the ordinary Pearson residual with respect to normal approximation, but the simulation results suggest that one cannot assume their distribution to be a standard normal distribution. Furthermore, when the Kolmogorov-Smirnov test was applied to the each dataset of size $n$ (sample size), instead of on each sample of size 10,000 of each individual residual, the empirical nominal level was very close to the theoretical nominal level thus suggesting the residuals $\breve{R}$ are indeed standard normally distributed. Such behavior was even better than those of the corrected Pearson residuals. Thus suggesting that to assess model adequacy one should use the PCA Pearson residual $\breve{R}_i$.

One should notice that the PCA Pearson residual $\breve{R}_i$ should not be the only residual used in a diagnostic analysis of an EFNLM since they are not suitable to identifying outlying observations. The reason is that they are obtained as linear combinations of the adjusted Pearson residuals and thus they ``lose'' the correspondence with the respective response variable. Therefore, our suggestion is that one uses both the PCA Pearson residuals $\breve{R}$ and the corrected Pearson residuals when conducting a diagnostic analysis on an EFNLM. Therefore, we expect these two residuals to become part of the essential toolkit of a practitioner that uses the exponential family nonlinear models to analyze data.

\section*{Acknowledgments} The authors would like to thank CNPq for their financial support.


\begin{thebibliography}{10}
\bibitem{1111} Cordeiro, G.M. (2004) On Pearson's residuals in generalized linear models, Statist. Prob. Lett. 66, 213-219.
\bibitem{cbvd}  Cordeiro, G.M., Paula, G.A. (1989) Improved likelihood ratio statistic for exponential family nonlinear models,
Biometrika 76, 93-100.
\bibitem{2} Cordeiro, G.M., Simas, A.B. (2009) The distribution of Pearson residuals in generalized linear models. Comp. Stat. Data Anal. 53, 3397-3411.
\bibitem{1aa} Cox, D.R., Hinkley, D.V. (1974) Theoretical Statistics, Chapman and Hall, London.
\bibitem{7} Cox, D.R., Snell, E.J. (1968) A general definition of residuals. J. R. Statist. Soc. B. 30, 248-275.
\bibitem{bb}  Fahrmeir, L., Kaufmann, H. (1985) Consistency and asymptotic normality of the maximum likelihood estimator in generalized linear models, Ann. Statist. 13, 342-368 (Fahrmeir and Kaufmann have some corrections for some of their results,Ann. Statist. 14, p. 1643).
\bibitem{aa}  Lehmann, E.L., Casella, E. (1998) Theory of Point Estimation, 2nd ed., Springer-Verlag, NewYork.
\bibitem{1} Loynes, R.M. (1969) On Cox and Snell's General Definition of Residuals. J. R. Statist. Soc. B. 31, 103-106.
\bibitem{as1} McCullagh, P., Nelder, J.A., (1989) Generalized Linear Models. Chapman and Hall, London.
\bibitem{aa123} Nelder, J.A., Wedderburn, R.W.M. (1972) Generalized linear models. J. Roy. Statist. Soc. A 135, 370-384.
\bibitem{33} Paula, G.A. (1992)  Bias correction for exponential family nonlinear models, J. Stat. Comput. Simul. 40, 43-54
\bibitem{12121} Pierce, D.A., Schafer, D.W. (1986) Residuals in generalized linear models. J. Amer. Statist. Assoc. 81, 977-986.
\bibitem{2222} Pregibon, D. (1981) Logistic regression diagnostics. Ann. Statist. 9, 705-724.
\bibitem{srs} Rocha, A.V., Simas, A.B. (2016) Independent and exactly distributed residuals for normal linear models based on principal component analysis. \emph{Submitted}. 
\bibitem{11} Simas, A.B., Cordeiro, G.M. (2009) Adjusted Pearson residuals in exponential family nonlinear models. J. Stat. Comp. Simul. 79, 411-425.
\bibitem{cc} Wei, B.-C. (1998) Exponential Family Nonlinear Models, Springer, Singapore.
\bibitem{1qa}  Williams, D.A. (1984) Residuals in generalized linear models, Proceedings of the 12th International Biometrics Conference,
Tokyo, 59-68.
\bibitem{135r34r} Williams, D.A. (1987) Generalized linear model diagnostics using the deviance and single case deletions, Appl. Stat. 36, 181-191.








\end{thebibliography}
\end{document}